\documentstyle[prl,aps,multicol]{revtex}
\input{epsf}

\begin{document}
\draft

\title{Sources and sinks separating domains of left- and right-traveling 
waves:\\ experiment versus amplitude equations } \author{
  Roberto Alvarez$^1$, Martin van Hecke$^{2*}$ and Wim van Saarloos$^2$}

\address{$^1$Department of
  Physics, Drexel University, 32$^{nd}$ and Chestnut Streets,
  Philadelphia, PA 19104, USA}
\address{$^2$Instituut--Lorentz, Leiden University, P.O. Box 9506,
  2300 RA Leiden, the Netherlands } 

\date{\today} \maketitle
\begin{abstract} 
  In many pattern forming systems that exhibit traveling waves,
  sources and sinks occur which separate patches of oppositely
  traveling waves. We show that simple qualitative features of their
  dynamics can be compared to predictions from coupled amplitude
  equations. In heated wire convection experiments, we find 
a  discrepancy between the observed multiplicity of sources and
  theoretical predictions.  The expression for 
  the observed motion of sinks is incompatible with any amplitude equation
  description.
\end{abstract}


\begin{multicols}{2}
  Since its inception \cite{newell}, the amplitude equation approach
  has grown out to become an important organizing principle of the
  theory of non-equilibrium pattern formation --- it has not only
  enabled us to uncover a number of general features of near-threshold
  pattern dynamics, but has also allowed us to understand the influence of
  boundary conditions, defects, etc. Many qualitative and quantitative
  predictions have been successfully confronted with experiments
  \cite{ch}. The most detailed comparison with experiments has been
  made for the type of systems for which the theory was originally
  developed, hydrodynamic systems that bifurcate to a stationary
  periodic pattern. For traveling wave systems, the range of validity
  of the appropriate amplitude equation is, however, much more an open
  question, both because the theoretical derivation has been performed
  for only a few systems \cite{schopf}, and because direct tests are
  difficult. Moreover, in practice complications often arise due to
  the presence of additional important slow variables \cite{coupling}.

  It is the aim of this paper to point out that sources and sinks that
  separate patches of traveling wave states, can provide a clear way
  of testing the consistency of the experimental observations with
  {\em generic qualitative} predictions from amplitude equations.
  Sources and sinks are distinguished by whether the group velocity
  points out- or inwards --- see Fig.\ 1. They occur in a wide
  variety of systems where oppositely traveling waves suppress each
  other --- in directional solidification \cite{bechhoefer}, the
  printer instability \cite{rabaud}, eutectic growth \cite{faivre},
  as well as  in convection \cite{dubois1,dubois2} --- but their
  properties have remained largely unexplored. We illustrate the idea
  we put forward with experiments on traveling waves occurring in a
  liquid heated by a wire just below the surface
  \cite{dubois1,dubois2,alvarez}. In the parameter range we have
  been able to explore (dimensionless control parameter
  $0.25\!\lesssim\!\varepsilon\!\lesssim\!0.5$), the experimental
  properties of sources and sinks we observe are {\em inconsistent}
  with the behavior predicted by the standard coupled amplitude
  equations for the near-threshold behavior in a one-dimensional
  system with left- and right-traveling waves \cite{ch}
\begin{mathletters}\label{coup1dCGL},
\begin{eqnarray}
  (\partial_{t} + &s_0 \partial_{x} )A_R =\varepsilon(1 + i c_{0})A_R
  + (1 + i c_{1}) \partial_{x}^{2}A_R -\nonumber \\ & (1 - i
  c_3)|A_R|^2 A_R - g_2 (1 - i c_2) |A_L|^2 A_R ~, \label{campa}\\ 
  (\partial_{t} -& s_0 \partial_{x} )A_L =\varepsilon(1 + i c_{0})A_L+
  (1 + i c_{1}) \partial_{x}^{2}A_L -\nonumber\\ & (1 - i c_3)|A_L|^2
  A_L - g_2 (1 - i c_2) |A_R|^2 A_L~.
\end{eqnarray} \end{mathletters}
In these equations, we have used suitable units of space and time, and
$A_R$ is the amplitude of the right traveling mode
$e^{-i(\omega_ct-k_cx)}$ and $A_L$ the one of the left traveling mode
$e^{-i(\omega_ct+k_cx)}$. Furthermore $\varepsilon$ is the
 control parameter which measures the distance from the
threshold of the instability at $\varepsilon\!=\!0$, and the
parameters $c_0 - c_3$ are related to the linear ($c_0$, $c_1$) and
nonlinear ($c_2$, $c_3$) dispersion of the waves. It is well-known
that, strictly speaking, the above equations only arise as the lowest
order amplitude equations if the linear group velocity $s_0$ is of
order $\varepsilon^{1/2}$; in practice, the equations are often also
applied to cases in which $s_0$ is finite at threshold, for lack of a
good alternative. This amounts to the idea that since
(\ref{coup1dCGL}) includes all the necessary terms and respects all
the proper symmetries, it is not unreasonable to hope that these
equations still provide a good qualitative description outside their
proper range of validity \cite{spanish}.

When the coupling parameter $g_2$ in (\ref{coup1dCGL}) is larger than
$1$, the left- and right-traveling waves suppress each-other
\cite{ch}, and the system evolves to a state consisting of patches
where either $A_L$ or $A_R$ is zero. Within such a patch, a {\em
  single} amplitude equation suffices, and the group velocity term
can be removed by a Galilean transformation. While many experimental and
theoretical studies have focused on this situation, we wish to
study the sources and sinks that connect these   patches of
left- and right-traveling waves (see Fig.\ 1). These coherent structures
involve both amplitudes $A_R$ and $A_L$ and therefore the group
velocity terms can not be removed. A study of their properties
may shed some light on the applicability of (\ref{coup1dCGL})
to real patterns. 

Our experimental set-up, shown in Fig.\ 2, is a simple system based on
a electrically heated wire immersed below the surface of a fluid
\cite{dubois1,dubois2,alvarez}. Beyond a critical heating
\begin{figure}
\epsfxsize=.6\hsize \vspace*{0.05 \hsize}

\hspace*{0.12 \hsize}  
(a)
\vspace{-0.28\hsize}

\mbox{\hspace*{0.16 \hsize} \epsffile{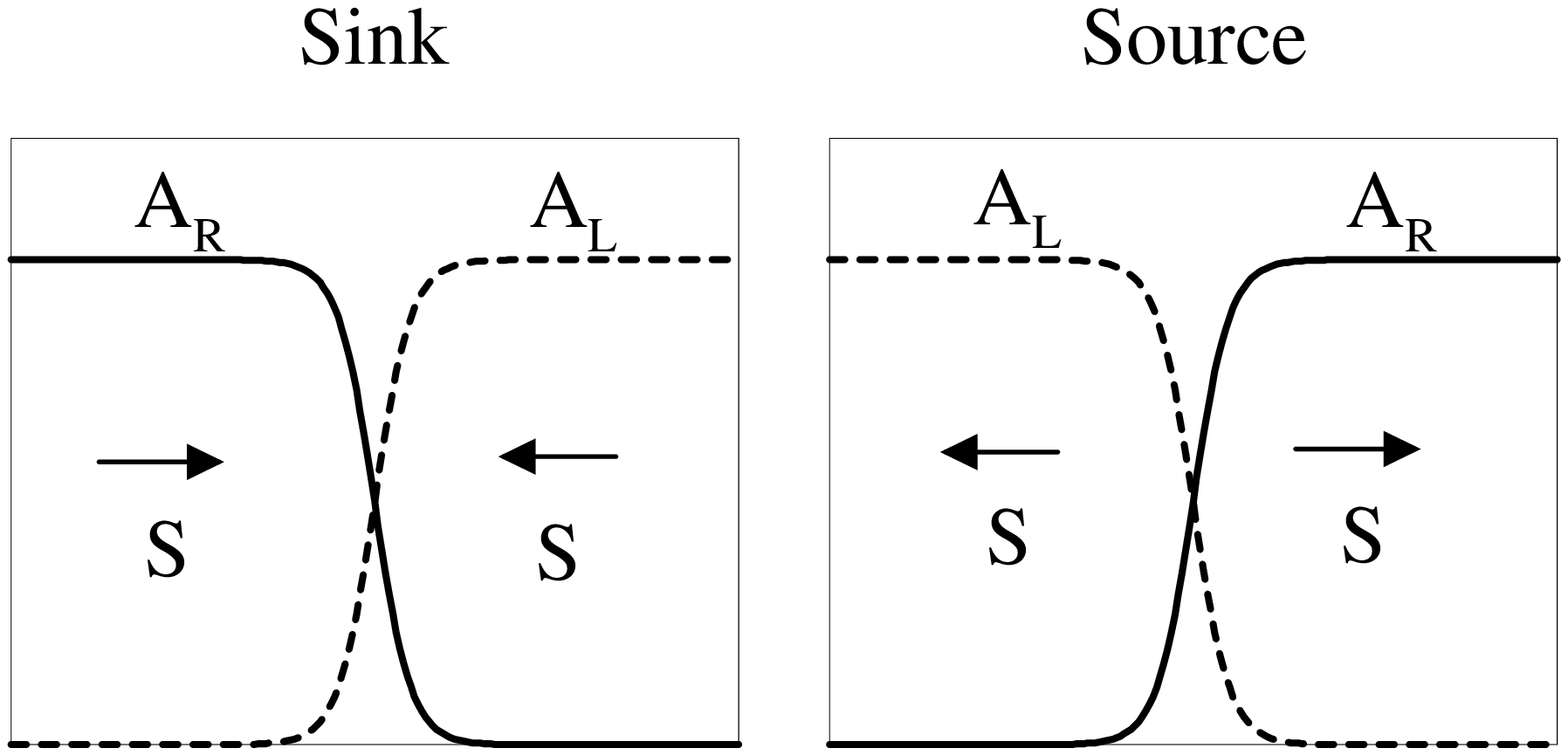}}\\

\epsfxsize=.5\hsize
\hspace*{0.12 \hsize} 
(b)
\vspace*{-0.15 \hsize}

\mbox{\hspace*{0.2 \hsize} \epsffile{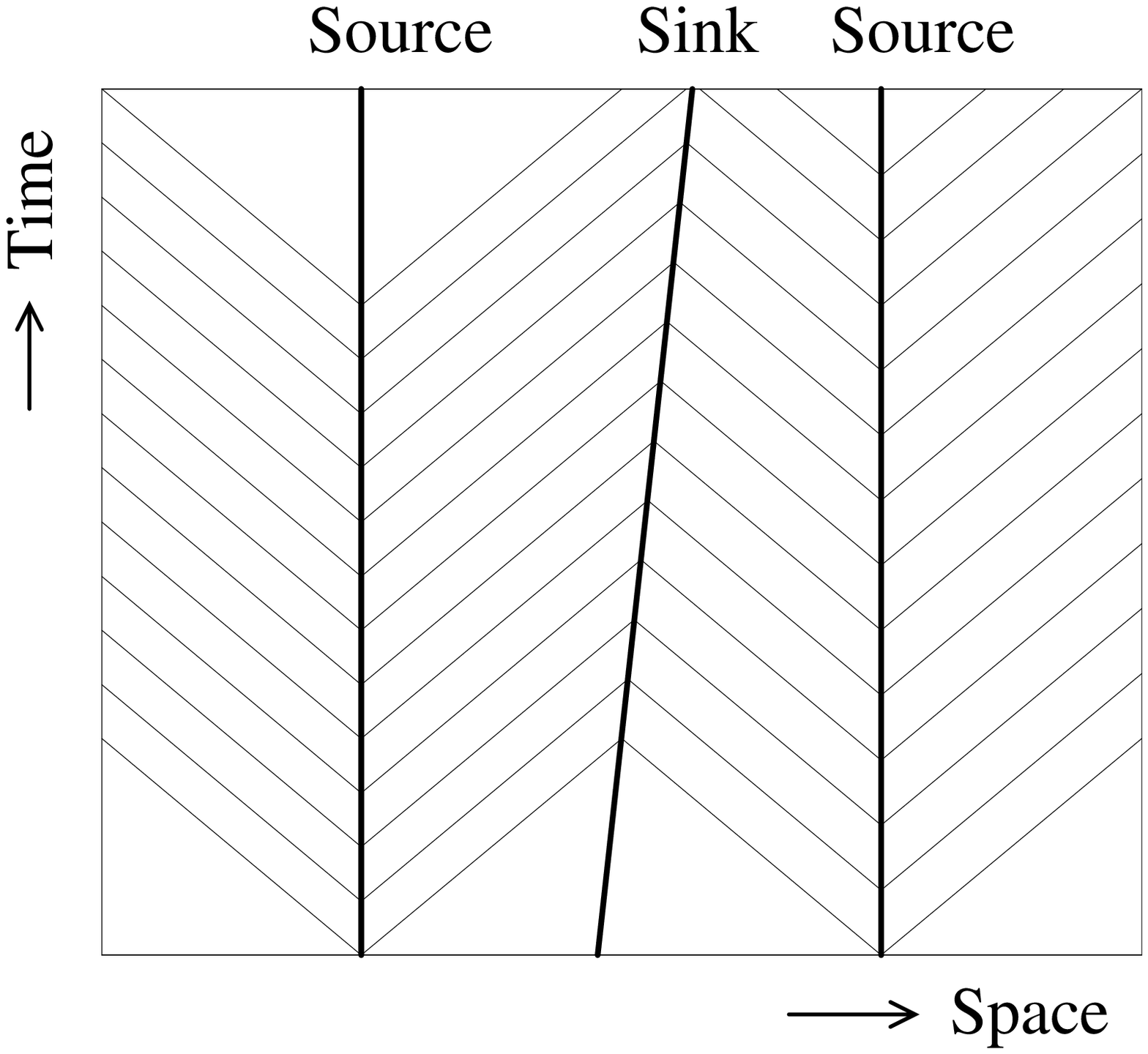}}
\vspace*{0.01 \hsize}
\narrowtext
\caption[] { Definition of sources and sinks. {\em
    (a)} Illustration of sources and sinks as coherent structures in
  terms of the behavior of the amplitudes $A_L$ and $A_R$ of 
  the left- and right
  traveling waves near these structures. A source is defined as a
  coherent structure at which waves with total group velocity $s$ pointing
  outward are generated, a sink one at which waves with  group velocity
  $s$ pointing inwards annihilate each other. {\em (b)}
  Illustration of the kinematics of sinks and sources in terms of the
  properties of the adjacent waves, for the case that the group
  velocity $s$ has the same sign as the phase velocity (as in the
  experiment, where $s\!\approx\!v_{ph}/3$). In {\em (b)},
  the definition of source and sinks given under {\em (a)} agrees with
  the intuitive notion that the waves travel away from a source and
  into a sink. In this figure, the thin  lines indicate lines of
  constant phase of the traveling waves. In accord with our
  experimental obervations, illustrated in Fig.\ 3b, we have
  drawn a case with two stationary and symmetric sources, each
  generating waves with different wavenumbers and frequencies, and one
  sink moving according to the phase matching rule. According to this
  rule, every constant phase line coming in at the source from the
  left   matches up 
  with a constant phase line coming in from the right: Phases at the
  sink match. A simple
  geometric construction gives the velocity of a sink at which phases
  match in terms of the frequencies and wavenumbers of the incoming
  waves --- see Eq.\ (\ref{phasematch}).  }
\end{figure}
\noindent 
power $Q_c$,
traveling periodic modulations appear at the surface via a forward
Hopf bifurcation \cite{dubois1,dubois2}.  The apparatus is similar to
the one used in \cite{dubois1,dubois2} but the design of the
Plexi-glass cell is somewhat different and larger (55x15x6 cm), so
that the sides are further away from the wire. Both top to bottom and
lateral views are possible in our set-up; in particular, the lateral
view proved especially useful for recording the time series signals. A
tungsten wire with a diameter of 0.1 mm is heated by means of an
electrical current and immersed in oil in the
middle of the cell and parallel to the longest side. The viscosity of
the GE SF 96 silicone oil is 0.5 stoke. Both the voltage
across the wire and the heating current were continuously monitored
to check that their values did not change during the measurements.
Four springs (two on each end of the cell) provide the necessary
tension to keep the wire parallel to the surface of the fluid
throughout the cell. A pair of micrometers attached to the ends of the
cell enables us to carefully adjust the wire-surface distance.  A
shadowgraph technique is used to record the signal generated by the
waves. The cell is illuminated with unpolarized white light. Two
photo-detectors can be placed at adjustable positions along the wire,
and the light signal captured by the detectors is sent to a digital
oscilloscope.  This temporal signal has a very local character thanks
to the diverging geometry from the light source towards the
acquisition plane. It therefore allows us to measure the local
frequency very accurately, even though the relation between the signal
and the surface modulations is  quite nonlinear due to the
strongly inhomogeneous temperature distribution in the direction
perpendicular to the wire. Since a single measurement may take several
hours due to the typical long times that fluids need to reach a steady
state condition, the temperature of the surrounding must be
controlled. For this reason, air conditioning was steadily supplied
and the temperature was continuously monitored.
\begin{figure}
\vspace*{-0.04 \hsize}
\epsfxsize=.8\hsize
\hspace*{0.06 \hsize}
\mbox{\epsffile{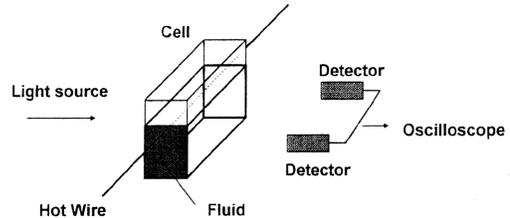}}
\narrowtext
\caption[]{Schematic drawing of the experimental setup. The
  frequencies of the traveling waves can be probed at two positions
  with photo-detectors. The distance between the surface of the liquid
  and the wire was varied from 1 to 3 mm.}
\end{figure}
The picture of the typical sequence of events can be drawn as follows
\cite{dubois1,dubois2,alvarez}.  Having chosen an adequate depth for
the wire, domains of left- and right-traveling waves emerge after the
power $Q$ is turned on. These patches are
separated by sinks and sources. Sources in our experiment send out waves
to both sides, while sinks have oppositely traveling waves coming in
from both sides.  Once transients have died out, the sources stay at
some fixed position while  the sinks generically  move,  either
towards a source (in
which case they usually annihilate each other) or a boundary (thus
also disappearing from the scene). A typical example is shown in Fig.\
3b. The time that a simple state, say
one with two or three sources and sinks, remains in the cell is
quite arbitrary; in the end a source usually is  the
longest living  object \cite{dubois2,alvarez}. We have mainly explored
the range $0.25\!\lesssim\!\varepsilon\!\lesssim\!0.5$ for the
 control parameter $\varepsilon\!\equiv\!Q/Q_c\!-\!1$. A
space-time plot of a source solution in this regime is shown in Fig.\ 
2 of \cite{dubois2}, while sideways snapshots of regions of the cell
with a source and a sink are shown in Figs.\ 3c and 3d.

Our main {\em experimental observations} concerning the dynamics of
sources and sinks are the following:\\ {\em (i)} The relative motion
of sinks and sources is independent of their separation, and so there
does not appear to be a long-range interaction between them.\\ {\em
  (ii)} Sources always have zero velocity, $v_{so}\!=\!0$, and are
symmetric: the wavenumber and frequency of the outcoming
left-traveling mode are always the same as those of the outcoming
right-traveling mode. The data that illustrate the stationarity of
the source are shown
 in Fig.\ 3b, while the fact that the waves emerging from a source are
 symmetric is illustrated by the photo-detector data shown in Fig.\
 3a. In this figure, the frequencies of the signals form the detectors
 D$_1$ and D$_2$ at both sides of
  the left source are exactly the same, and so are those of the
  signals taken at opposite sides of the
  source  on the right by detectors D$_3$ and D$_4$.\\ {\em (iii)}
While sources are  stationary and symmetric, they are not
unique: each source sends out waves with a well-defined frequency
and wave number, but different sources send out different waves
--- compare, e.g., the two sources of Fig. 3a: the frequency of the
signals of D$_1$ and D$_2$ is different from that of D$_3$ and D$_4$. 
We take this as evidence that in these experiments at
least a one-parameter family of sources exists.\\ {\em (iv)} 
As Fig.\ 3b illustrates, sinks typically
move. Moreover, most of our sinks are found to move in such a way that
the incoming phases {\em match} at the sink \cite{note1}: in the frame
traveling with the sink, the frequencies of the waves
coming in from both sides are equal and no phase difference builds
up across these sinks. This was already illustrated in Fig.\ 1b.
If we write the two appropriate incoming modes
as $e^{-i(\omega_Rt-ik_Rx)}$ and $e^{-i(\omega_Lt+ik_Lx)}$, then the
velocity $v_{si}^{match}$ of such a sink is immediately found to be
\begin{equation}
\label{phasematch}
v_{si}^{match} = \frac{\omega_R-\omega_L}{k_R+k_L}~.
\end{equation}
This relation implies that when  a sink is sandwiched between two
sources, it moves away from the source with the largest frequency
and its velocity is completely determined by the properties of
the adjacent  sources.

We now confront these results with {\em theoretical predictions}.
Since the pattern occurs via a forward Hopf bifurcation 
\cite{dubois1,dubois2}, the generic
 amplitude equations are given by Eqs. (\ref{coup1dCGL}).
We take $g_2\!>\!1$ since
traveling wave states occur, and since the group velocity 
is is about a third of the phase velocity in this
experiment,  we allow the linear group velocity $s_0$ to be of order $1$.
Note that for our analysis, we do not need knowledge of the values of
the other parameters occuring in (1) \cite{note2}.

Property {\em (i)}, the absence of long-range interactions between
sources and sinks, is consistent with the fact that amplitudes at both
sides of source and sink solutions approach their asymptotic value
exponentially fast, as in the single-mode equation, Eq.$\,$(\ref{campa}) with $A_L\!=\!0$ \cite{physd,Bohr}.

\begin{figure} 
\epsfxsize=.85\hsize
\hspace*{0.08 \hsize}
\mbox{\epsffile{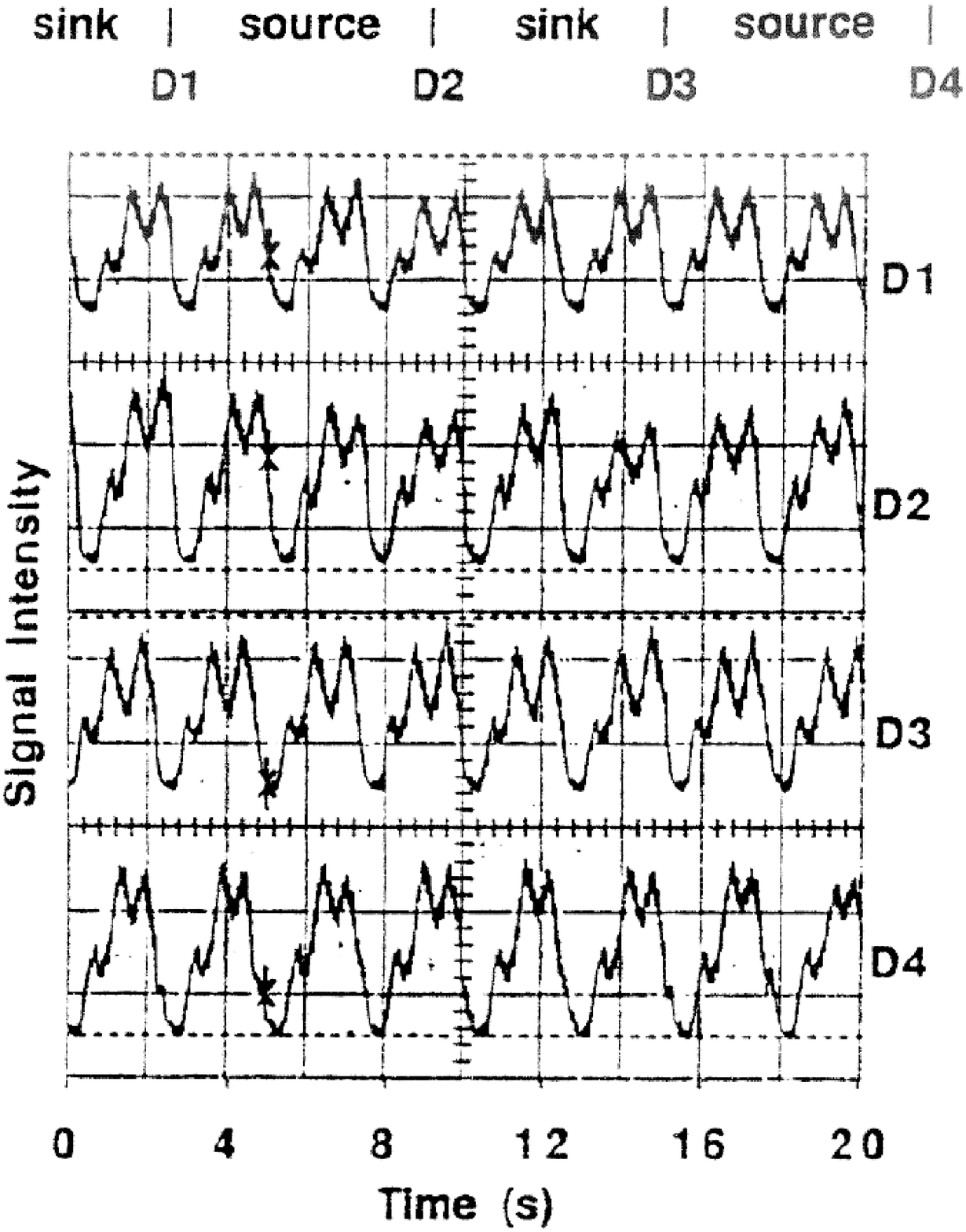}}\\
\vspace*{-0.6 \hsize}

\hspace*{0.02 \hsize} (a) \vspace*{0.55 \hsize}

\hspace*{0.02 \hsize} (b)
\vspace*{-0.12 \hsize}

\epsfxsize=.75\hsize
\hspace*{0.11 \hsize} 
\mbox{\epsffile{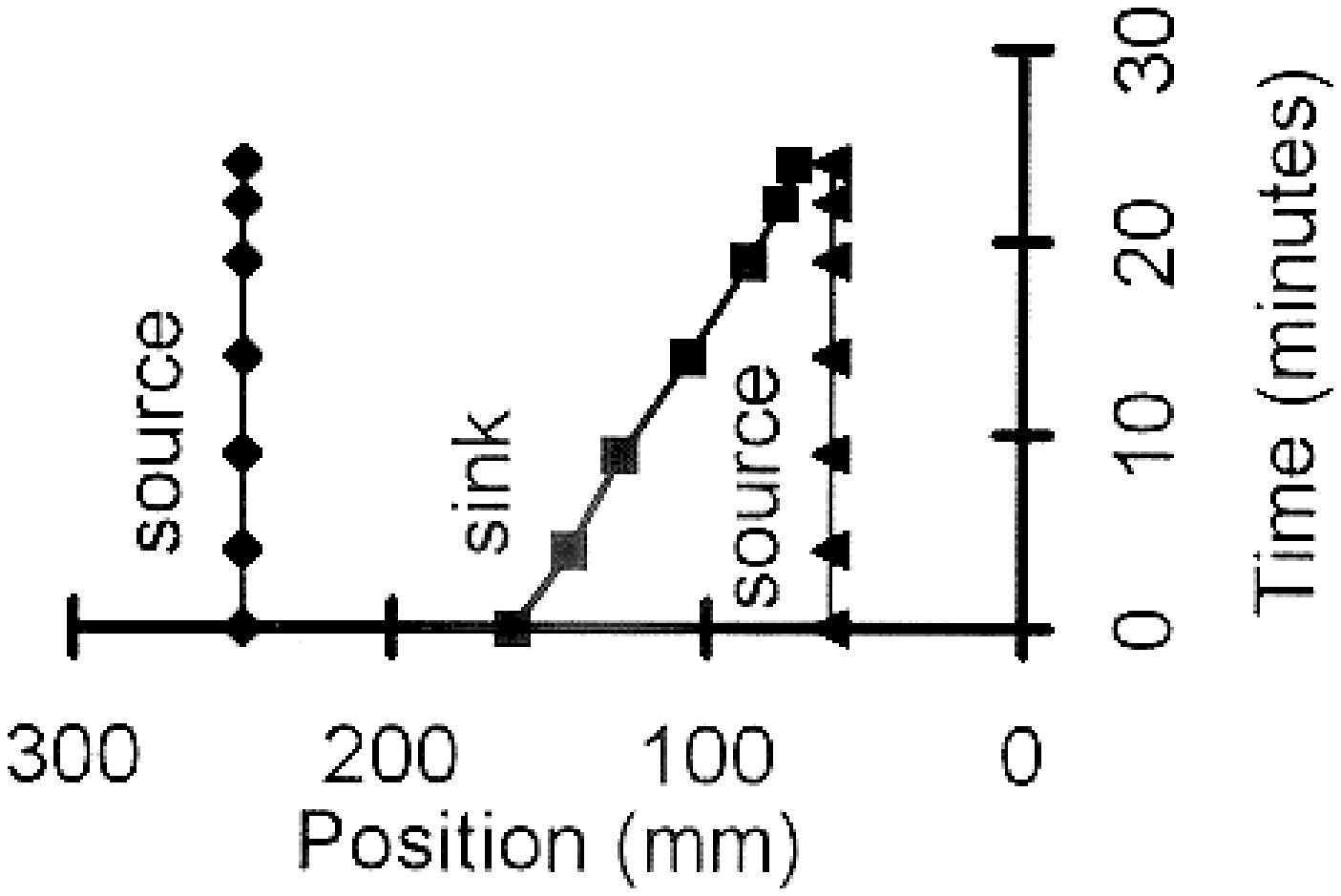}}\\
\vspace*{0.02 \hsize}

\epsfxsize = .80 \hsize 
\hspace*{0.02 \hsize} (c) \vspace*{-0.06 \hsize}

\hspace*{0.08 \hsize}
\epsffile{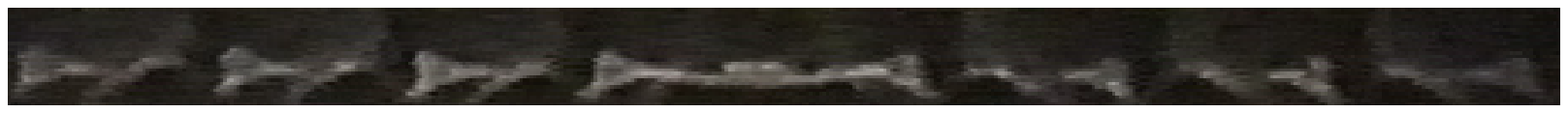}\\ 
\vspace*{0.01 \hsize}

\epsfxsize = 0.80 \hsize
\hspace*{0.02 \hsize} (d) \vspace*{-0.06 \hsize}

\hspace*{0.08 \hsize}
\epsffile{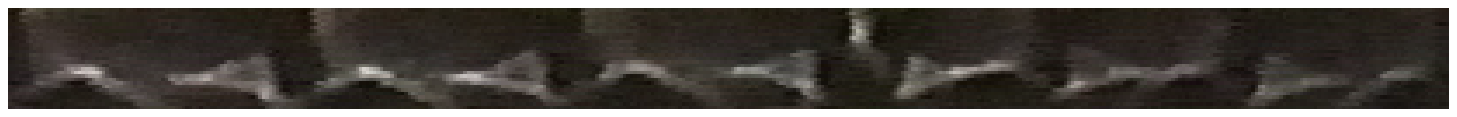}
\vspace*{0.05 \hsize}

\narrowtext
\caption[]{Experimental results.
  {\em (a)} Example of the signals from the photodetectors
  placed at various positions in between source and sink solutions, as
  indicated.  Typical signal amplitudes are about a factor 8 above the
  noise level. The absolute value of each signal is arbitrary, especially the
  frequency is relevant.
  Note that the frequencies of the two waves sent out by
  each separate source are exactly the same, but that the two sources
  send out different waves. As a result, the sink is sandwiched in
  between different incoming waves. {\em (b)} Example of traces of the
  position of sinks and sources in the experiment. Compare Fig. 1b,
  where a similar situation is drawn schematically.
   {\em (c,d)}  Snapshots of
  regions of the experimental cell with a source ({\em c}) and a sink
  ({\em d}), taken from a sideways video image of the cell. Note
  the asymmetry of the pattern around the source, which is roughly in
  the middle of {\em (c)}, and the asymmetry of the pattern
  around the sink, which is slightly right of center in {\em (d)}.}
 \end{figure}

To compare with observations {\em (ii)} and {\em (iii)}, we have
analyzed the generic existence and multiplicity of {\em source}
solutions of (\ref{coup1dCGL}) with an extension of previous counting
arguments\cite{physd} for solutions of the form
$A_R=e^{-i\omega_0t}\hat{A}_R(x-v_{so}t)$ and
$A_L=e^{-i\omega_0t}\hat{A}_L(x+v_{so}t)$. Our analysis
\cite{counting} shows that independent of the specific values of the
parameters, source solutions of (\ref{coup1dCGL}) {\em generically}
come in discrete sets. In particular, one typically expects there to
be only a {\em unique} symmetric source solution with $v_{so}\!=\!0$,
and numerical simulations of (\ref{coup1dCGL}) confirm this.  This is
in clear contradiction with the experiments, where we find a
continuous family of them!

We now turn to {\em sinks}, which according to {\em (iv)} move in the
experiments with a velocity (\ref{phasematch}) \cite{note3}.
Can the
phase matching
property of the sinks underlying this equation be reproduced at all in
an amplitude approach based on (\ref{coup1dCGL})?  The answer is {\em
  no}. To see this, note that $v_{si}^{match}$ is according to
(\ref{phasematch}) given in terms of the {\em total} frequencies
$\omega_R$, $\omega_L$ and wave numbers $k_R$, $k_L$ of the incoming
modes. In an amplitude expansion, these are written as an expansion
about their critical values, e.g.,
$\omega_R\!=\!\omega_c+\omega_{A_R}$ where $\omega_{A_R}$ is the
frequency of the amplitude $A_R$ of the right-traveling mode, etc. In
terms of these, the experimentally observed velocity of phase-matching
sinks becomes
\begin{equation}
\label{phasematch2}
v_{si}^{match} =
\frac{\omega_{A_R}-\omega_{A_L}}{2k_c+k_{A_R}+k_{A_L}}~,
\end{equation}
which underscores once more the fact that this velocity depends on
both the fast and the slow spatial scales. However, as an amplitude
description is based on an adiabatic elimination of the fast scales,
the amplitude equations (\ref{coup1dCGL}) do not involve the
parameters $\omega_c$ and $k_c$ associated with the fast scales. So,
although families of moving sink solutions do exist for
(\ref{coup1dCGL}), there is no mechanism in these equations to single
out the velocity (\ref{phasematch2}) as the selected velocity of
sinks.

In summary, our results demonstrate that generic properties of sources
and sinks in traveling wave systems, like their multiplicity and
dynamics, allow a simple yet powerful comparison between experiments
and amplitude equation descriptions. For the heated wire convection
experiment in the range $0.25\!\lesssim\!\varepsilon\!\lesssim\!0.5$
our experiments are inconsistent with an amplitude description.
Although a final conclusion must await further study of the
$\varepsilon\!\rightarrow\!0$ limit, our results point to
two important issues. First of all, they question the soundness of
using Eqs.\ (\ref{coup1dCGL}) for systems with finite group velocity
$s_0$. Secondly, they provide a clear example of the possible
importance of non-adiabatic effects (coupling of the slow and fast
scales \cite{nonadi}) in sinks, even though the two are widely
separated, as the relative frequency modulation $\Delta
\omega/\omega_c$ (which is comparable to the ratio of the typical sink
velocity and the phase velocity) can be as small as 1/50 in our
experiments.

 R.A. is grateful to N. Kwasnjuk for help in constructing the
 experimental cell.

\end{multicols}


\begin{references}
\bibitem{note2} Although we don't need this in the analysis, the
  stability of the traveling waves necessitates to take $c_1c_3\!<\!1$
  \cite{ch}.
\bibitem{note3} When the phases match across a sink, one might be
  tempted to say that no phase-slips occur there.  We prefer to avoid
  to use the word phase-slip in this context, however, as it usually
  refers to events at which the amplitude goes through zero in a
  single amplitude equation\cite{physd}. In the case of two coupled
  modes, each separate amplitude does not need to exhibit
  phase slip events even if the phases across a sink do not match.
  The relation between phase-slips of the amplitudes and of the
  patterns is quite different for single and double amplitude
  equations.
\end{references}
\end{document}